\documentclass{article}
\usepackage{spconf,amsmath,graphicx}

\usepackage{enumitem}

\setlist{nosep, leftmargin=14pt}

\usepackage{mwe} 
\usepackage{color} 

\usepackage{subcaption}
\usepackage{url}

\title{Quantifying white matter hyperintensity and brain volumes\\ in heterogeneous clinical and low-field portable MRI}
\name{%
\begin{tabular}{@{}c@{}}
Pablo Laso $^{1,2}$ and 
Stefano Cerri $^{1,3}$ and 
Annabel Sorby-Adams $^{1}$\quad
Jennifer Guo $^{1}$\quad
Farrah Mateen $^1$\\
\textit{Philipp Goebl $^{4}$\quad
Jiaming Wu $^{4}$\quad
Peirong Liu $^{1}$\quad
Hongwei Li $^{1}$\quad
Sean I. Young $^{1}$\qquad 
Benjamin Billot $^{5}$}\\
\textit{Oula Puonti $^{1,6}$\quad
Gordon Sze $^{7}$\quad
Sam Payabavash $^{7}$\quad
Adam DeHavenon $^{7}$\quad
Kevin N. Sheth $^{7}$}\\
\textit{Matthew S. Rosen $^{1}$\quad
John Kirsch $^{1}$\quad
Nicola  Strisciuglio $^{2}$ \quad
Jelmer  M. Wolterink} $^{2}$ \\
\textit{Arman Eshaghi $^{4}$\quad
Frederik Barkhof $^{4}$ \quad
W. Taylor Kimberly $^{1}$\quad
Juan Eugenio Iglesias $^{1,4,5}$}
\end{tabular}}

\address{$^1$ Massachusetts General Hospital, Harvard Medical School, USA \\
$^2$ Electrical Engineering, Mathematics and Computer Science, University of Twente, The Netherlands \\
$^3$ Pioneer Centre for AI, University of Copenhagen, Denmark \\
$^4$ Centre for Medical Image Computing, University College London, UK \\
$^5$ Computer Science and Artificial Intelligence Laboratory, Massachusetts Institute of Technology, USA\\
$^6$ Danish Research Centre for Magnetic Resonance, Denmark \quad $^7$ Yale School of Medicine, USA}

\begin{document}
%
\maketitle

\begin{abstract}

Brain atrophy and white matter hyperintensity (WMH) are critical neuroimaging features for ascertaining brain injury in cerebrovascular disease and   multiple sclerosis. 
Automated segmentation and quantification is desirable but existing methods require high-resolution MRI with good signal-to-noise ratio (SNR). This precludes application to clinical and low-field portable MRI (pMRI) scans,  thus hampering  large-scale tracking of atrophy and WMH progression, especially in underserved areas where pMRI has huge potential.
Here we present a method that segments white matter hyperintensity and 36 brain regions from scans of any resolution and contrast (including pMRI) without retraining.  We show results on eight public datasets and on a private dataset with paired high-  and low-field scans (3T and 64mT), where we attain strong correlation between the WMH ($\rho$=.85) and hippocampal volumes ($\rho$=.89) estimated at both fields. Our method is publicly available as part of  FreeSurfer, at: {http://surfer.nmr.mgh.harvard.edu/fswiki/WMH-SynthSeg}.

\end{abstract}
%
%
\section{Introduction}
\label{sec:intro}
White matter hyperintensity (WMH) on magnetic resonance imaging of the human brain is  associated with stroke, cognitive decline, and cardiovascular disease. WMH is frequently detected in brain MRI scans in the general population with chronic disease such as hypertension. A recent observational study was performed in a safety net emergency setting evaluating  adult patients with a vascular risk factor who were being evaluated  for a non-stroke complaint. In this cohort, more than half of the subjects had WMH identified on portable, low field MRI~\cite{deHavenon2023}. In addition, WMH is a hallmark of multiple sclerosis (MS), a disease that creates a demyelination process that may lead to disability~\cite{dobson2019multiple}. The MS disease process is correlated with other neurodegeneration, leading to abnormally high atrophy rates in different brain regions~\cite{fisher2008gray}. Closer monitoring of WMH and atrophy is thus desirable at a larger scale.

Inexpensive portable MRI (pMRI) technology is becoming increasingly available for imaging WMH in the community at large scale. For example, the low-field (64mT) Swoop system (Hyperfine Inc) produces images that agree well with  high-field counterparts when WMH are scored by a radiologist~\cite{deHavenon2023}. A crucial component of large-scale deployment is automated segmentation and quantification of WMH and brain regions, as manual identification and tracing of regions of interest (ROIs) in 3D is impractical and irreproducible. 

Quantification of WMH and brain anatomy (including atrophy) is also very desirable in clinical MRI. As opposed to a research MRI, which is typically isotropic, clinical scans often comprise fewer slices acquired in 2D. These take less time for clinical review and are  less susceptible to motion artifacts. Precise quantitative analysis of these scans would allow closer tracking of atrophy and WMH progression.

A large array of methods exist for segmenting brain anatomy and WMH. Representative classical methods include: FreeSurfer~\cite{fischl2002whole} and FSL~\cite{patenaude2011bayesian}
for brain ROIs;  LST~\cite{schmidt2012lst} and BIANCA~\cite{GRIFFANTI2016191}
for WMH; or SAMSEG~\cite{puonti2016fast,cerri2021contrast},  which segments both. Machine learning techniques, often using convolutional neural networks (CNNs), include: QuickNat~\cite{roy2019quicknat} or FastSurfer~\cite{henschel2020fastsurfer}, for brain ROIs; or \cite{brosch2016deep,ghafoorian2017location} for WMH.
These  methods are designed for conventional high-field MRI (1.5-3T), and often have requirements in terms of resolution (typically 1mm isotropic), pulse sequence (often T1-weighted for anatomy, FLAIR for WMH), or both. Therefore, they struggle with the huge variability in orientation (axial, coronal, sagittal), resolution, and contrast of clinical MRI in real scenarios. This problem is exacerbated in pMRI, where the low field imposes  limitations in signal-to-noise ratio (SNR) that are compensated with large voxel sizes, and where the geometry of the scanner often leads to severe signal loss away from its center. While domain adaptation~\cite{wang2018deep} can mitigate these problems to some extent, a CNN than can handle any MRI contrast and resolution without retraining is highly desirable.

Here we present WMH-SynthSeg, a CNN that segments WMH and brain anatomy from scans of any resolution and contrast, including low-field pMRI. WMH-SynthSeg builds on our previous work on domain randomization~\cite{BILLOT2023102789,synthsr_science}
to achieve such agnosticity. Compared with our previous method for simultaneous segmentation of WMH and anatomy~\cite{billot2021joint}, WMH-SynthSeg: \textit{(i)}~does not require retraining; \textit{(ii)}~uses a specific WMH model and a composite loss to improve sensitivity and specificity; \textit{(iii)}~adapts to low-field MRI; and \textit{(iv)}~uses multi-task learning for enhanced robustness. We show that, as a result, WMH-SynthSeg can robustly segment WMH and anatomy from clinical and pMRI.

\vspace{-1mm}
\section{Methods}
\label{methodology}
\vspace{-1mm}
\subsection{Synthetic training data}
\label{data generation}
\vspace{-1mm}
WMH-SynthSeg relies on a synthetic MRI generator similar to~\cite{synthsr_science}, which requires a training dataset with $N$ 1mm isotropic T1-weighted (T1w) scans $\{I_n\}$ and corresponding 3D segmentations $\{S_n\}$; these are defined on the same 1mm isotropic grid and include labels for brain ROIs and WMH. 

At every iteration during training: \textit{(i)}~a random pair ${(I_n, S_n)}$ is selected; \textit{(ii)}~$(I_n, S_n)$ are augmented non-linear deformation; \textit{(iii)}~a Gaussian mixture model conditioned on the labels is sampled independently at every voxel, with means and variances that are randomly sampled from uniform distributions -- except for the WMH class (details below); \textit{(iv)}~the Gaussian image is corrupted by a random smooth bias field; \textit{(v)}~random orientation and resolution are simulated (via smoothing) to synthesize a lower resolution scan; and \textit{(vi)}~the low-resolution scan is upsampled to the original 1mm isotropic grid. This process generates: the upsampled synthetic scan $I^{syn}$, deformed segmentation $S$, deformed real image $I$, and bias field $B$. All these are defined on the original 1mm grid (see~\cite{synthsr_science} for examples of synthetic images). 

The generator has 4 key improvements compared with~\cite{synthsr_science}:

\noindent\textit{(i)} The mean intensity of the WMH class is not distributed across the whole range 0-255. Instead, we simulate WMH in T2-like  sequences (including FLAIR) and WM hypointensity in T1w-like sequences. This is done as follows: when the white matter (WM) mean is high (over 128), we constrain the WMH mean to be lower than the WM mean (T1w-like). Conversely, when the WM mean is below 128, we constrain the WMH mean to be greater than the WM mean (T2-like). 

\noindent\textit{(ii)} The standard deviation of the noise (Gaussian variances) and bias field strength is twice as large as in~\cite{synthsr_science}, to accommodate the lower SNR and stronger signal losses of pMRI. 

\noindent\textit{(iii)} The generator produces not only $I^{syn}$ but also a deformed image $I$ and a bias field $B$ that will be used as regression targets by the CNN in a multi-task learning setting. This boosts the robustness of the CNN as shown in the experiments.

\noindent\textit{(iv)} The sampling scheme for the random resolution covers a wider spectrum of acquisitions. 25\% of the time, we generate 1mm isotropic images, to support high-resolution scans. Another 25\% we generate clinical scans of random orientation with 1mm in-plane resolution and random slice spacing between 2.5mm and 8.5mm. 25\% of the scans mimic the resolution of the stock sequences that the Hyperfine Swoop ships with (axial with $\sim$1.5mm in plane and 5mm spacing). The final 25\% simulates more isotropic scans acquired at low field, with random voxel sizes between 2-5 mm in every direction.

\vspace{-1mm}
\subsection{Model architecture and training}
\vspace{-1mm}
WMH-SynthSeg uses a 3D U-net~\cite{ronneberger2015unet}
with five levels, 64 feature maps per level, and group normalization~\cite{wu2018group}. Each level has two convolutions (kernel size: 3x3x3) followed by ReLU activations.
The final layer has $L+2$ channels: the first $L$ correspond to the labels and are fed to a softmax layer to produce soft segmentations; the last two correspond to the predicted bias field and high-resolution T1w intensities.

Training uses the Adam optimizer
to minimize a loss function consisting of four terms with equal weight: the cross-entropy and Dice scores between the predicted and ground truth segmentations; the average $\ell_1$ error of the predicted T1w intensities (normalized such that the median intensity of the WM is 1); and the $\ell_1$ error of the predicted bias field (in logarithmic scale):
\vspace{-1mm}
$$
\mathcal{L} = CE(S, \hat{S}) - AvDice(S, \hat{S})) + |I-\hat{I}|  + |\log B- \log \hat{B}|,  
$$
where $\hat{S}$, $\hat{I}$, and $\hat{B}$ are the predictions for the segmentation, T1w intensities, and bias field, respectively.

We note that, while training with Dice may be more common in segmentation, combining it with cross-entropy has two advantages. First, it provides a more informative gradient in the first iterations of training, when the gradient of the Dice loss is rather flat. And second, it explicitly penalizes false positives in scans without WMH -- in which the Dice score for the WMH is zero independently of the prediction. In addition, including $I$ and $B$ in the loss increases the robustness of the method, as shown by the experiments below.


At test time, the input scan is resampled to 1mm isotropic resolution and fed to the CNN. Test-time augmentation is performed by left-right flipping the image, flipping the output back, and averaging with the non-flipped version.  The first $L$ channels of the output yield the final segmentation; the outputs corresponding to the bias field and the T1w intensities are a potentially useful by-product, but are disregarded here.

We train the CNN with PyTorch using 160$^3$ voxel patches. The validation loss typically converges in $\sim$10$^5$ iterations.

\section{EXPERIMENTS AND RESULTS}

\subsection{Datasets}

We used nine different datasets in our experiments, some just for training (``Tr''), some for testing (``Te''), and some for both using cross validation (``Tr/Te'').

\noindent\textbf{HCP}~\cite{van2013wu} (Tr): 897 1mm isotropic scans of young subjects from the Human Connectome Project. We used FreeSurfer to automatically segment the anatomy into 36 ROIs.

\noindent\textbf{ADNI}~\cite{jack2008alzheimer} (Tr): 1148 1mm isotropic scans from the ADNI. We used FreeSurfer to segment the anatomy and WMH.

\noindent\textbf{GE3T} (Tr/Te): 20 cases with 1mm isotropic T1w and 1x1x3mm axial FLAIRs. This a subset of the WMH segmentation challenge~\cite{kuijf2019standardized}. We combined the automated FreeSurfer segmentation of the T1w with the manual delineations available for the FLAIRs into a single ground truth segmentation.

\noindent\textbf{Singapore} (Tr/Te): another subset of the  challenge with 20 cases from a separate site (same MRI acquisitions and labels).  

\noindent\textbf{Utrecht} (Tr/Te):another subset with 20 cases from a third site.

\noindent\textbf{ISBI}~\cite{carass2017longitudinal} (Tr/Te): 15 1mm isotropic T1w scans (segmented with FreeSurfer) and 1x1x2mm axial FLAIRs with manually traced WMH (merged with the anatomy into one label map).    

\noindent\textbf{FLI-IAM}~\cite{commowick2018objective} (Tr/Te): T1w and FLAIR scans from 15 cases with varying resolution but all close to 1mm isotropic. Consensus WMH tracings are available from 7 raters, which we merged with the FreeSurfer segmentations of the T1w scans. 

\noindent\textbf{ADHD}~\cite{bellec2017neuro} (Te): 20 1mm isotropic T1w scans from typically developing control children and adolescents and no WMH. 

\noindent\textbf{MGH} (Te): 12 MS patients 
from our hospital (MGH) with 1mm  T1w and FLAIR, as well as pMRI axial T1w and FLAIR (in-plane resolution: 1.6-1.8mm; slice spacing: 5-6mm).

\subsection{Competing methods}
\label{competing methods}

We compare our method with: \textit{(i)}~SAMSEG~\cite{puonti2016fast,cerri2021contrast}, which is a Bayesian method that is adaptive to MRI contrast, and is (to our best knowledge) the only existing method that can readily segment anatomy and WMH from scans acquired with any pulse sequence; and \textit{(ii)}~LST-LPA~\cite{schmidt2017bayesian}, which yields great performance on FLAIR  acquisitions but does not work on other MRI contrasts. We also consider two ablations of our method to assess the importance of its components: a version with just Dice in the loss (similar to~\cite{billot2021joint} but with domain randomization), and a version without the prior on the mean of the WMH class.
We note that LST and SAMSEG operate at the native resolution of the scan, whereas WMH-SynthSeg always produces a 1mm isotropic segmentation.

\subsection{Experimental setup}

We analyze the performance of our proposed method WMH-SynthSeg with three different experiments. The first experiment assesses the performance of the method directly with Dice scores. We first trained WMH-SynthSeg using GE3T and Singapore (using 15 scans for validation), and tested on ISBI, FLI-IAM, and Utrecht. We then reversed the roles to obtain Dice scores for GE3T and Singapore. We note that HCP and ADNI were also part of the training dataset in both folds. We note that training inputs are all synthetic and that the real images  are only used as regression targets.

The second experiment assesses false positive rates (FPR) using young healthy controls from the ADHD dataset. Since WMH is not expected in these scans, we can use the estimated  WMH loads as a proxy for  FPR. The model in this experiment is trained with all the datasets from the first experiment. 

The third experiment assesses the ability of the methods to segment  pMRI data, using the same model as in the second experiment. We used the FreeSurfer segmentations of the high-field 1mm T1w scans as ground truth for the anatomy, and the LST segmentations of the high-field 1mm FLAIRs as ground truth for the WMH. Since accurate co-registration of low- and high-field scans is difficult due to nonlinear geometric distortions, we use the correlation between the ground truth and estimated ROI volumes to assess performance.

\subsection{Results}

Table~\ref{tab:results_comparison_DICE} shows the average Dice across the high-field datasets in the first experiment, for the WMH and for 23 representative brain ROIs: brainstem,  and left/right cortex, WM, hippocampus, amygdala, thalamus, caudate, pallidum, putamen, accumbens, and cerebellum cortex and WM (we exclude less reliable ROIs, e.g., accumbens). WMH-SynthSeg outperforms the competing methods across the board. 
The ablations show that cross-entropy and multi-task learning  have a moderate positive impact on the segmentation of anatomy, whereas the prior on mean of WMH component greatly boosts the performance of the WMH segmentation. In absolute terms, our new method yields competitive Dice scores for anatomy (Dice=.85 for isotropic T1w) and WMH (Dice=.62 in FLAIR, higher than SAMSEG and LST). We also highlight its capability to produce useful WMH segmentations from the T1w, with Dice scores as high as those of the competing methods in FLAIR.

\begin{figure*}[th!]
    \centering
        \includegraphics[width=.965\linewidth]{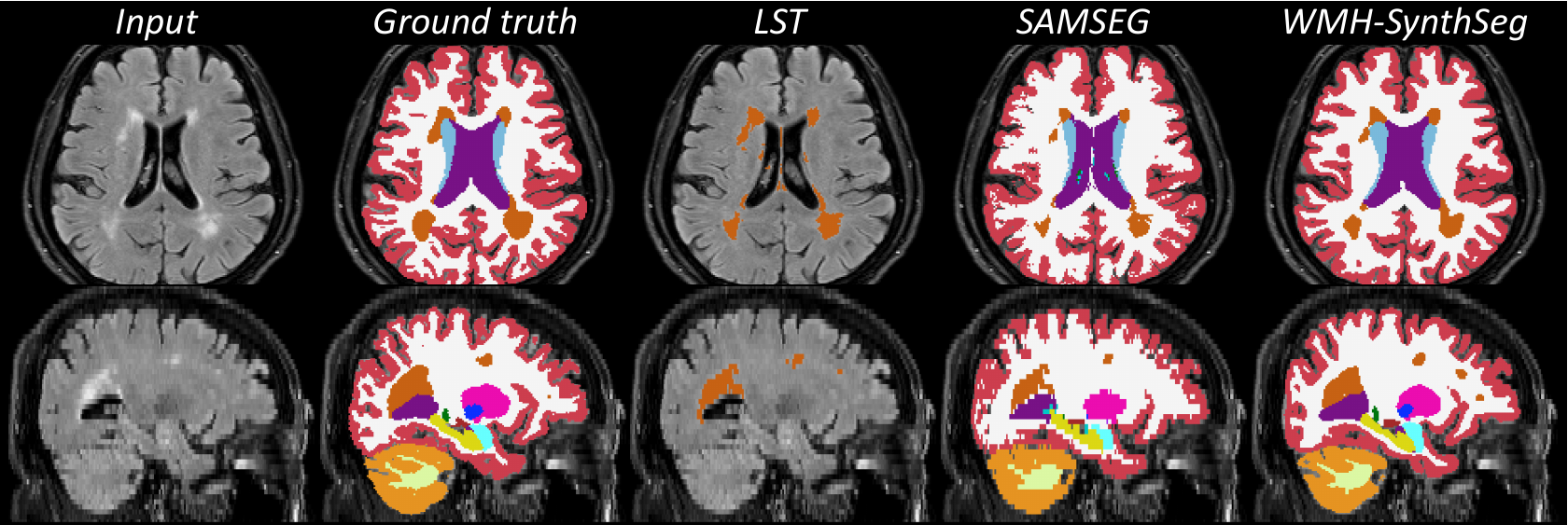}
    \caption{Input, ground truth, and automated segmentations of a sample high-field scan from the Singapore dataset. The top row shows the high-resolution axial view; the bottom row shows a lower resolution orthogonal view (in sagittal orientation).}
    \label{fig:examples}
\end{figure*}

\begin{table}[!hb]
\begin{small}
    \begin{tabular}{|c|c|c|c|c|} \hline 
        \textbf{Method} & \multicolumn{2}{c|}{\textbf{T1w}} & \multicolumn{2}{c|}{\textbf{FLAIR}} \\ \cline{2-3} \cline{4-5}
         & \textbf{Anat} & \textbf{WMH} & \textbf{Anat} & \textbf{WMH} \\ \hline 
        LST (LPA) & N/A & N/A & N/A & 0.57 \\ \hline 
        SAMSEG & 0.81 & 0.46 & 0.72 & 0.56 \\ \hline 
         \begin{tabular}{@{}c@{}}WMH-SynthSeg \\ (NoWMH-noCE-noMTL) \end{tabular}  & 0.83 & 0.47 & 0.76 & 0.53\\ \hline
        WMH-SynthSeg (NoWMH)        & \textbf{0.85} & 0.47 & 0.78 & 0.54 \\ \hline
        WMH-SynthSeg (full)  & \textbf{0.85} & \textbf{0.55} & \textbf{0.79} & \textbf{0.62} \\ \hline
    \end{tabular}
    \centering
    \caption{Average Dice scores for anatomy (averaged over 23 ROIs) and WMH, on high-field T1w and FLAIR scans. NoWMH-noCE-noMTL is the ablation without prior on the WMH mean, cross-entropy term in the loss, or multi-task learning (i.e., similar to~\cite{billot2021joint}). NoWMH is the ablation without the prior on the mean of the WMH intensities.}
    \label{tab:results_comparison_DICE}
    \end{small}
\end{table}

Figure~\ref{fig:examples} shows a qualitative comparison on a FLAIR scan from the Singapore dataset, both in the high-resolution axial plane, and in a lower resolution orthogonal view (sagittal). LST produces crisp segmentations of the WMH at native resolution, but with many false positives around the septum pellucidum (between the ventricles). SAMSEG, which also operates at native resolution, struggles with partial voluming (e.g., for the cortex) and often undersegments WMH. Our method, on the other hand, produces \emph{isotropic} segmentations that are accurate for both anatomy and WMH.

\begin{table}[!t]
\begin{small}
    \begin{tabular}{|c|c|c|c|c|} \hline 
        \textbf{Method} & \multicolumn{2}{c|}{\textbf{T1w}} & \multicolumn{2}{c|}{\textbf{FLAIR}} \\ \cline{2-3} \cline{4-5}
         & \textbf{Hippo} & \textbf{WMH} & \textbf{Hippo} & \textbf{WMH} \\ \hline 
        LST (LPA) & N/A & N/A & N/A & -0.33 \\ \hline 
        SAMSEG & 0.71 & 0.63 & 0.69 & 0.64 \\ \hline 
        WMH-SynthSeg (full)  & \textbf{0.89} & \textbf{0.75} & \textbf{0.86}  & \textbf{0.85} \\ \hline
    \end{tabular}
    \centering
    \caption{Correlation between ground truth volumetric measurements obtained from high-field (FreeSurfer from T1w for anatomy, LST from FLAIR for WMH) and from automated segmentations of the pMRI (MGH dataset). The hippocampal volumes (``Hippo'') are left-right averaged. }
    \label{tab:low_field}
    \end{small}
\end{table}

In the FPR experiment with young controls, our method produces on average 950~mm$^3$. This is a low value comparable to that produced by SAMSEG (877~mm$^3$); we note that LST is not compatible with the ADHD dataset as it has T1w contrast. The ablated versions show increases to 1,150~mm$^3$ (without the WM mean prior) and 1,850~mm$^3$ (without the prior or multi-task learning), highlighting the contribution of these components to the accuracy of the algorithm. 

\begin{figure}[!t]
    \centering
        \includegraphics[width=\linewidth]{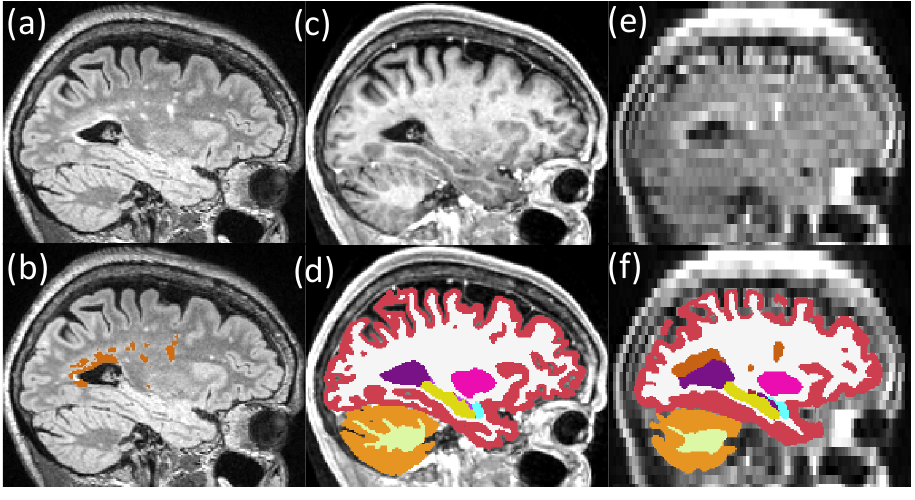}
    \caption{(a)~High-field 1mm isotropic FLAIR from MGH dataset. (b)~LST segmentation, used as ground truth for WMH. (c)~High-field 1mm T1w. (d)~FreeSurfer segmentation of (c), used for ground truth for anatomy. (e)~pMRI of the same subject at 2x2x5.8mm axial resolution. (f)~WMH-SynthSeg segmentation. We note that, despite affine alignment of the high-field images to the pMRI, the anatomy on the slices is slightly different due to nonlinear distortion.}
    \label{fig:pMRI}
\end{figure}

Finally, Table~\ref{tab:low_field} shows the correlations between the volumetric measurements derived from the high-field scans (ground truth) and the pMRI, for the WMH and for a representative brain ROI (the hippocampus, which is tightly connected with aging and many brain diseases, e.g., dementias). LST completely fails at low field, as it was not designed for it. Being contrast agnostic, SAMSEG yields fairly strong correlations (between .63 and .71). WMH-SynthSeg produces very strong correlations (12-21 points higher than SAMSEG). This is attributed to its excellent ability to adapt to low-field images, which is qualitatively exemplified in Figure~\ref{fig:pMRI}.

\section{Conclusion}

We have presented the first method that can simultaneously segment brain ROIs and WMH in scans of any resolution and contrast, including  pMRI. Future work will include  realistic modeling of WMH and evaluation on pMRI from larger cohorts. WMH-SynthSeg is publicly available and has potential in analyzing pMRI  acquired in medically underserved areas.


\section{ACKNOWLEDGMENTS}
Supported by a grant from the Jack Satter Foundation and by NIH grants RF1MH123195, R01AG070988, R01EB031114, UM1MH130981, RF1AG080371, and R01NS112161.

\bibliographystyle{IEEEbib}
\bibliography{strings,refs}

\end{document}